# تصحیح تابشی انرژی کازیمیر برای میدان اسکالر با شرط مرزی مخلوط در ۳+۱ بعد

مددعلی ولوئیان*،۱

۱گروه فیزیک، واحد سمنان، دانشگاه آزاد اسلامی، سمنان، ایران

## چکیده

در این مقاله، مرتبه صفرم و مرتبه اول از تصحیح تابشی انرژی کازیمیر برای میدان اسکالر (جرم دار و بدون جرم) محدود شده با شرط مرزی مخلوط (دیریکله-نیومن) بین دو صفحه موازی در ۳+۱ بعد محاسبه شده است. دو نکته در فرایند انجام این محاسبه قابل توجه است. یکی از این نکات، استفاده از یک برنامه بازبهنجارش متفاوت و البته سازگار با شرایط مرزی حاکم بر مساله است. در این برنامه بازبهنجارش جهت دستیابی به پارامترهای فیزیکی موجود در لاگرانژی (به عنوان مثال: جرم میدان و ثابت جفت شدگی) از کانترترمهای سازگار با شرط مرزی و البته وابسته به مکان استفاده می شود. نکته دیگر در فرایند این محاسبه، استفاده از روش منظم سازی کم کردن جعبه ها است. در روش منظم سازی کم کردن جعبه ها، مشابه با ساختار اصلی یک ساختار دیگر در نظر گرفته می شود و اختلاف انرژی های خلا مربوط به این دو ساختار در حدهای مناسب محاسبه می شود. این امر موجب می شود واگرایی های دخیل در محاسبات بنحو شفاف و روشنی یکدیگر را حذف نمایند. پاسخ های بدست آمده برای هر دو مرتبه انرژی کازیمیر (مرتبه صفر و مرتبه اول از تصحیح تابشی) در نمودارهای مختلف رسم شده و این نمودارها نشان می دهند که سازگاری مناسب و فیزیکی در پاسخها وجود دارد. نکته حائز اهمیت اینکه پاسخ بدست آمده در این مقاله در مرتبه اول تصحیح تابشی با پاسخ گزارش شده در این مرتبه متفاوت است اما سازگاری های قابل انتظار فیزیکی را بر آورده می کند.

**کلیدواژگان:** انرژی کازیمیر، بازبهنجارش، شرط مرزی، منظم سازی

## مقدمه

بیش از ۶۰ سال پیش انرژی کازیمیر به عنوان یک اثر ماکروسکوپیک ناشی از قطبش خلاء[۱] توسط کازیمیر[۲] مطرح گردید[۱]. مقاله کازیمیر برای مدتی ناشناخته باقی ماند، تا اینکه اولین تلاش برای مشاهدهٔ این پدیده توسط اسپارنای[۳] در سال ۱۹۵۸ انجام گرفت[۲].

هرچند دقت آزمایش صورت گرفته توسط اسپارنای بسیار بالا نبود، اما اندازه گیری های دیگری پس از او و با برطرف نمودن نقایص آزمایش اسپارنای انجام شد، که بر درستی پیش بینی کازیمیر صحه گذاشتند. همانطور که میدانیم انرژی کازیمیر از اختلاف بین دو

* نویسنده مسئول: m-valuyan@sbu.ac.ir

---

1 Vacuum polarization
2 H.B.G. Casimir
3 M.J. Sparnaay

انرژی خلاء[۱] در صورت حضور شرط مرزی غیر بدیهی[۲] و عدم حضور آن حاصل می‌شود. اثر کازیمیر را می‌توان به عنوان یکی از جالب ترین تجلیات ویژگی‌های غیر بدیهی حالت خلاء در نظریه کوانتومی میدان[۳] برشمرد که در واقع ناشی از تغییر دو قطبش خلاء توسط شرایط مرزی یا هندسه است. به عبارت دیگر، این اثر نتیجه‌ای از اعوجاج و دگرش در طیف نوسانات خلاء با محدود شدن حجم کوانتش و یا غیر بدیهی شدن توپو لوژی[۴] فضا است. همانطور که می دانیم مجموع انرژی ناشی از فرکانس‌های مجاز در خلاء، یک مقدار واگرا[۵] و بینهایت است. وقتی در همین خلاء یک شرط مرزی دلخواه را قرار می دهیم فرکانس‌های مجاز تغییر خواهند کرد؛ اما باز هم مجموع انرژی ناشی از کل این فرکانسها واگراست. کازیمیر در سال ۱۹۴۸ نشان داد که اختلاف این دو مقدار واگرا (انرژی خلاء) منجر به یک کمیت مشاهده پذیر[۶]، قابل اندازه گیری و البته متناهی[۷] خواهد شد که بعدها به نام انرژی کازیمیر شهرت یافت [۳-۵]. تصحیحات تابشی[۸] انرژی کازیمیر اولین بار توسط برداگ[۹] و همکاران ایشان در سال ۱۹۸۵ مورد بررسی واقع گردید، و در سالهای بعد نیز توجه ویژه ای به این مقوله صورت پذیرفته است[۶]؛ بطوریکه برای میدانهای کوانتومی مختلف با شرایط مرزی گوناگون در ساختارهای[۱۰] متعدد این کمیت محاسبه شده است. آن چیزی که در محاسبه تصحیحات تابشی انرژی کازیمیر اهمیت دارد، بازبهنجارش پارامترهای برهنه[۱۱] موجود در لاگرانژی است (به عنوان مثال: جرم میدان[۱۲] و یا ثابت جفت شدگی[۱۳]). همانطور که می‌دانیم

وظیفه بازبهنجارش[۱۴] پارامترهای برهنه موجود در لاگرانژی معمولا برعهده کانترترمها[۱۵] نهاده می‌شود و این کانترترمها هستند که در برنامه بازبهنجارش واگرایی‌های ناشی از پارامترهای برهنه را در خود هضم کرده و پارامترهای فیزیکی لاگرانژی را بدست می دهند. در اجرای برنامه بازبهنجارش در بسیاری از مقالات گذشته از کانترترمهای فضای آزاد[۱۶] برای بازبهنجارش کمیتهای برهنه استفاده شده است [۷-۱۰]. این امر حتی در مسائلی که میدان کوانتومی در آنها تحت اثر یک شرط مرزی بوده است، نیز انجام پذیرفته است. عقیده ما براین است که، اگر در مساله ای میدانهای کوانتومی تحت اثر یک شرط مرزی قرار دارند، بنابراین برنامه بازبهنجارش مربوط به این میدان نیز باید بگونه‌ای اجرا شود که با شرط مرزی اعمال شده، سازگاری مناسب را داشته باشد. در واقع انتظار می‌رود که، همه اثرات ناشی از شرط مرزی روی تک تک المانهای مربوط به برنامه بازبهنجارش تاثیرگذار باشد. در این امر، کانترترمها به عنوان یکی از المانهای برنامه بازبهنجارش نمی توانند مستثنا باشند. لذا، استفاده از کانترترم مربوط به فضای آزاد (کانترترم خلاء) در برنامه بازبهنجارش، بدون در نظر گرفتن شرایط جدید حاکم برمیدان کوانتومی (مثلا شرط مرزی و یا توپولوژی فضا) نمی تواند درست باشد. عدم انتخاب صحیح کانترترم منجر به این خواهد شد که، برنامه بازبهنجارش بدرستی عمل نکرده و در نهایت برای تعدادی از کمیتهای فیزیکی (به عنوان مثال انرژی کازیمیر) مقادیر واگرا بدست آید [۱۱]. البته اعمال

---

شرایط مرزی حاکم بر مساله در برنامه بازبهنجارش منجر به بدست آمدن کانترترمهای وابسته به مکان[1]در آن خواهد شد. در تعدادی از مقالات گذشته این امر مورد توجه قرار گرفته است و برای اولین بار سادات گوشه و همکاران وی به معرفی یک روش بازبهنجارش پرداخته اند؛ که در آن شرط مرزی حاکم بر مساله در برنامه بازبهنجارش به صورت سیستماتیک[2] و سازگار[3] ورود یافته و این ورود منجر به یافتن کانترترمهای وابسته به مکان در برنامه بازبهنجارش شده است [۱۲]. با استفاده از این روش بازبهنجارش، تصحیحات تابشی انرژی کازیمیر برای میدان اسکالر با شرایط مرزی دیریکله[4] در یک، دو و سه بعد فضایی بین دو صفحه موازی بدست آمده است و پاسخهای بدست آمده در کلیه موارد نیز سازگار با مبانی شناخته شده فیزیکی بوده است [۱۵-۱۳]. ارائه یک پاسخ همگرا[5] برای تصحیح تابشی انرژی کازیمیر مربوط به میدان اسکالر در دو بعد فضایی، هم برای یک رویه تخت و هم برای یک رویه خمیده، که معمولا همراه با پیچیدگی های بیشتری است، نیز یکی دیگر از مزیتهای روش بازبهنجارش مذکور است [۱۶،۱۷]. در این مقاله، نیز با استفاده از روش بازبهنجارش مذکور، تصحیح تابشی انرژی کازیمیر را برای میدان اسکالر تعریف شده در نظریه $\phi^4$ بین دو صفحه موازی با شرط مرزی مخلوط[6] (دیریکله-نیومن) محاسبه خواهیم کرد. منظور از شرط مرزی مخلوط، شرط مرزی است که در آن تابع میدان کوانتومی باید روی یکی از صفحات شرط مرزی دیریکله را و روی صفحه مقابل شرط مرزی نیومن را ارضا نماید. شایان ذکر است که پاسخ های بدست آمده

در این مقاله با آنچه در [۱۱] ذکر شده است متفاوت بوده اما با مبانی قابل انتظار فیزیکی کاملاً سازگار است. در مرجع [۱۱] محاسبه تصحیح تابشی انرژی کازیمیر برای میدان اسکالر محدود شده با شرط مرزیهای مختلف در ابعاد یک، دو و سه بعد فضایی با استفاده از کانترترم آزاد انجام پذیرفته است. نتیجه این محاسبه برای ابعاد یک و سه بعد فضایی همگرا بوده است. اما پاسخ همین مساله در دوبعد فضایی واگرا بدست آمده است. در واقع به نظر می رسد کانترترم آزاد جهت بازبهنجارش پارامترهای برهنهی لاگرانژی در یک سیستم محدود شده با شرط مرزی غیربدیهی، قادر به حذف همه واگراییها نیست و همین امر منجر به بروز واگرایی در بعضی کمیتهای فیزیکی(به مانند انرژی کازیمیر) گردیده است. زیرا وقتی در همان مسائل بجای استفاده از کانترترم آزاد از کانترترم سازگار با شرایط مرزی استفاده می کنیم، نه تنها مساله را به صورت خودسازگار[7] حل کرده ایم، بلکه پاسخها در تمامی ابعاد فضایی اعم از بعدهای زوج و یا غیر آن همگرا و سازگار با مبانی فیزیکی بدست آمده است؛ و این همان چیزی است که در این مقاله نیز برای شرط مرزی مخلوط در ۳+۱ بعد انجام پذیرفته است.

جهت حذف واگراییهای موجود در فرایند محاسبات انرژی کازیمیر، استفاده از روشهای منظم سازی[8] اجتناب‌ناپذیر است. روش های مختلفی برای منظم سازی در طی شصت سال گذشته تاکنون معرفی شده است، که می توان از بعضی از آنها به روش تابع زتا[9]، روش تابع گرین[10] و روش بسط پراکندگی[11]، تکنیک

---

منظم سازی تابع قطع[۱] و منظم سازی ابعادی[۲] و... نام برد [۱۸–۲۰]. هریک از تکنیک‌های منظم‌سازی با توجه به مزیت ها و یا معایب خود در مسائل مختلف بکارگیری می‌شوند. در این مقاله از روش منظم سازی کم کردن جعبه ها[۳] استفاده شده است. برای محاسبه انرژی کازیمیر با استفاده از این روش، معمولا دو ساختار مشابه در نظر گرفته می شود، و انرژی های خلا این دو ساختار از هم کم می شوند. به عنوان مثال، برای محاسبه انرژی کازیمیر بین دو صفحه موازی با فاصله صفحات $a$ ، آنرا در بین دو صفحه  موازی دیگر  با فاصله صفحات $a > L$ محبوس می کنیم. نام این

ساختار را به مانند آنچه در شکل (۱) نشان داده شده است، با عنوان ساختار $A$ در نظر می‌گیریم. ساختار $B$ را نیز به مانند ساختار $A$ معرفی می کنیم. در ساختار $B$ نیز دو صفحه موازی به فاصله $b$ بین دو صفحه موازی دیگر به فاصله $b > L$ محبوس شده‌اند. حال رابطه انرژی کازیمیر برای دو صفحه موازی به فاصله $a$ بصورت زیر تعریف می شود:[۲۱،۲۲]

$$E_{\text{Cas.}} = \lim_{b \to \infty} \lim_{L \to \infty} (E_A - E_B)$$

(۱)

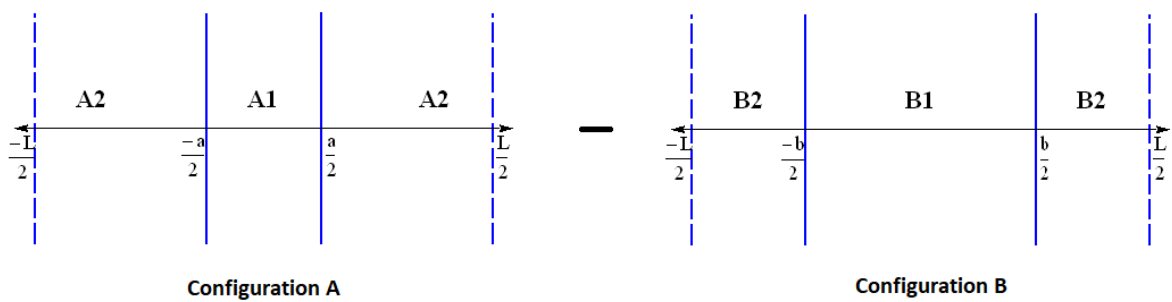

**Configuration A**     **Configuration B**

A2   A1   A2     B2   B1   B2

**شکل ۱.** ساختار رسم شده در سمت راست ساختار $B$ و ساختار رسم شده در سمت چپ ساختار $A$ را نشان می دهد. در هر دو ساختار دو صفحه موازی در بین دو صفحه موازی دیگر به فاصله $L$ محبوس شده اند.

که در آن $E_A$ و $E_B$ به ترتیب کل انرژی خلا مربوط به ساختارهای $A$ و $B$ است. در تعاریف قدیمی‌تر برای محاسبه انرژی کازیمیر، اختلاف انرژی خلا بدون حضور شرط مرزی(فضای مینکوفسکی[۴]) از انرژی خلا با حضور شرط مرزی محاسبه می‌شود. اما در روش کم کردن جعبه ها، فضای مینکوفسکی به ساختاری به مانند ساختار $B$ جایگزین شده است. به راحتی می توان نشان داد، که در حدهای مناسب ( $L \to \infty$ و $b \to \infty$) ساختار $B$ به مانند فضای مینکوفسکی عمل می‌کند. مزیت نسبی روش کم کردن جعبه ها این است که، پارامترهایی

که ناشی از ساختار $B$ بوده و در فرایند محاسبات ورود خواهند یافت به مانند منظم ساز[۵] عمل کرده و این منظم‌سازها به روند حذف واگرایی‌ها کمک شایانی می نماید. نکته دیگر در استفاده از روش کم کردن جعبه ها، بی نیاز بودن آن در استفاده از تمدید تحلیلی[۶] جهت حذف واگرایی‌ها است. همانطور که می‌دانیم، تمدید تحلیلی در محاسبات انرژی کازیمیر خود موجب فراهم شدن بستری برای بعضی از ابهامات می گردد؛ که بی نیاز بودن روش کم کردن  جعبه ها در استفاده از تکنیک تمدید تحلیلی به شفافیت آن در فرایند حذف واگرایی‌ها می‌افزاید[۲۱]. در حالت

---

کلی در منظم سازی کم کردن جعبه ها انرژی خلا دو ساختار مشابه در یک شرایط حدی خاص از هم کم می شوند. این شرایط حدی در عمل باید بگونه ای باشد تا، با تعریف انرژی کازیمیر – که در واقع اختلاف انرژی نقطه صفر یک ساختار با شرایط مرزی غیربدیهی از انرژی خلا مربوط به فضای مینکوفسکی است- منافاتی نداشته باشد. ما اعتقاد داریم که هرگاه بتوان چنین شرایط حدی را در اختلاف دو انرژی خلا مربوط به دو ساختار مشابه ایجاد کرد، می توان انرژی کازیمیر را با استفاده از روش کم کردن جعبه ها پیاده سازی نمود. در مقاله [۱۶,۱۷] این کار برای کره $S^2$ و $S^3$ صورت پذیرفته است و تعمیم آن برای دیگر فضاهای خمیده و یا نااقلیدسی نیز دور از انتظار نیست و می تواند ممکن باش. لذا در این مقاله نیز، جهت محاسبه مرتبه صفرم و اول از تصحیح تابشی انرژی کازیمیر برای میدان اسکالر با شرط مرزی مخلوط(دیریکله–نیومن) بین دو صفحه موازی در ۳+۱ بعد و برای حذف واگراییها از روش مذکور بهره جسته-ایم. پاسخهای بدست آمده در مرتبه صفرم انرژی کازیمیر با جوابهای موجود در گذشته تطابق کامل داشته و برای مرتبه اول تصحیح تابشی استفاده از روش مذکور موجب دستیابی به پاسخ-های متفاوت، امّا سازگار و فیزیکی گردیده است [۱۱]. در بخش آتی، در ابتدا به محاسبه مرتبه صفرم از انرژی کازیمیر برای میدان اسکالر با شرط مرزی مخلوط(دیریکله–نیومن) پرداخته، و در بخش سوم مرتبه اول تصحیح تابشی انرژی کازیمیر برای شرط مرزی مذکور محاسبه می‌شود. در بخش ۴ به جمع بندی نتایج حاصله می‌پردازیم.

## مرتبه صفر انرژی کازیمیر

رابطه انرژی نقطه صفر[1] خلا را برای میدان اسکالر محدود شده با شرط مرزی مخلوط بین دو صفحه موازی به فاصله $a$ از هم در ۳+۱ بعد می توان بصورت زیر نوشت:

$$E^{(0)} = \frac{1}{2} \iint \frac{L^2 d^2 k}{(2\pi)^2} \sum_{n=0}^{\infty} \omega_n$$

(۲)

که در آن فرکانس‌های $\omega_n = \sqrt{k_x^2 + k_y^2 + \frac{\pi^2}{a^2}\left(n+\frac{1}{2}\right)^2 + m^2}$ مجاز در انرژی خلا بوده و $m$ جرم مربوط به میدان اسکالر می باشد. برای محاسبه انرژی کازیمیر، قبل به بکارگیری تعریف ارائه شده در رابطه (۱) جهت به انجام رساندن روش کم کردن جعبه ها، از رابطه ابیل – پلانا به فرم ذیل برای تبدیل عبارت جمع[2] موجود در معادله (۲) به انتگرال استفاده می کنیم [۲۳]:

$$\sum_{n=0}^{\infty} \mathcal{F}\left(n+\frac{1}{2}\right) = \int_0^{\infty} \mathcal{F}(x)dx - i \int_0^{\infty} \frac{\mathcal{F}(it) - \mathcal{F}(-it)}{e^{2\pi t}+1}dt$$

(۳)

اولین جمله در سمت راست تساوی فوق معمولاً دارای مقداری واگراست و جمله انتگرالی[3] نامیده می‌شود و آخرین جمله در سمت راست تساوی فوق جمله برنچکات[4] گفته می شود که معمولاً دارای مقداری متناهی است. اکنون با استفاده از روابط (۱) و (۲) و همچنین با استفاده از رابطه فوق، می توان اختلاف انرژی های خلا دو ساختار $A$ و $B$ را بصورت زیر نوشت:

$$E_A^{(0)} - E_B^{(0)}$$
$$= \frac{L^2}{4\pi} \int_0^{\infty} kdk \left[\frac{a}{\pi}\int_0^{\infty}(k^2+\xi^2+m^2)^{\frac{1}{2}}d\xi \right.$$
$$\left. + \frac{2a}{\pi}\int_{\sqrt{k^2+m^2}}^{\infty}\frac{(\eta^2-k^2-m^2)^{\frac{1}{2}}}{e^{2a\eta}+1}d\eta\right]$$
$$+ \frac{L^2}{2\pi}\int_0^{\infty}kdk \left[\frac{L-a}{2\pi}\int_0^{\infty}(k^2+\xi^2+m^2)^{\frac{1}{2}}d\xi \right.$$
$$\left. + \frac{L-a}{\pi}\int_{\sqrt{k^2+m^2}}^{\infty}\frac{(\eta^2-k^2-m^2)^{\frac{1}{2}}}{e^{(L-a)\eta}+1}d\eta\right] - \{a\to b\}$$

---

1 Zero point energy
2 Summation term
3 Integral term
4 Branch-cut term

(٤)

جملات اول هر کروشه در عبارت فوق، که همان جمله انتگرالی رابطه ایپل-پلانا هستند، واگرا بوده و می توان نشان داد که این واگرایی به ازای کلیه مقادیر متناهی از $L$ ، $a$ ، $b$ و $m$ بصورت زیر حذف خواهند شد:

$$\frac{L^2}{4\pi^2}\left[a + 2\frac{L-a}{2} - b + 2\frac{L-b}{2}\right]\int_0^\infty kdk \int_0^\infty (k^2+\xi^2+m^2)^{\frac{1}{2}}d\xi = 0$$

(٥)

تنها جملات باقیمانده از رابطه (٤) همان جملات برنچکات ایپل-پلانا است و داریم: [١٤]

$$E_A^{(0)} - E_B^{(0)} = \frac{L^2 a}{2\pi^2}\int_m^\infty d\eta \int_0^{\sqrt{\eta^2-m^2}} kdk \frac{\sqrt{\eta^2-k^2-m^2}}{e^{2a\eta}+1} + \frac{L^2(L-a)}{2\pi^2}\int_m^\infty d\eta \int_0^{\sqrt{\eta^2-m^2}} kdk \frac{\sqrt{\eta^2-k^2-m^2}}{e^{(L-a)\eta}+1} - \{a\to b\}$$

(٦)

در گام آخر محاسبه انرژی کازیمیر، طبق تعریف ارائه شده در رابطه (١)، به محاسبه حد $L\to\infty$ و $b\to\infty$ می پردازیم. می توان نشان داد که کلیه جملاتی که در رابطه (٦) نوشته شده و مربوط به ناحیه‌های A2، B1 و B2 هستند، در این حد صفر خواهند شد و مرتبه صفرم از انرژی کازیمیر مربوط به میدان اسکالر جرمدار بین دو صفحه موازی با شرط مرزی مخلوط (دیریکله-نیومن) به صورت زیر بدست می آید:

$$E_{Cas.}^{(0)} = \frac{L^2 m^2}{8\pi^2 a}\sum_{j=1}^\infty \frac{(-1)^{j+1} K_2(2maj)}{j^2}$$

(٧)

که در آن $K_2(\alpha)$ تابع بسل اصلاح[1] شده است. معمولاً در مباحث مربوط به انرژی کازیمیر، بررسی حد بدون جرم[2] میدان و حد میدان با جرمهای سنگین[3] از مقوله های مورد توجه است. معادله (٧) برای میدانهای بدون جرم ($m\to 0$) و میدانی با جرمهای بسیار بزرگ ($m\gg 1$) به عبارت زیر منتهی خواهد شد: [١٤-١٧]

$$E_{Cas.}^{(0)} \to \begin{cases} \dfrac{7}{8}\dfrac{L^2\pi^2}{1440a^3} & as \quad m\to 0; \\[2ex] \dfrac{1}{16\pi a}m\sqrt{\dfrac{m}{\pi a}}\,e^{-2ma} & as \quad ma\gg 1. \end{cases}$$

(٨)

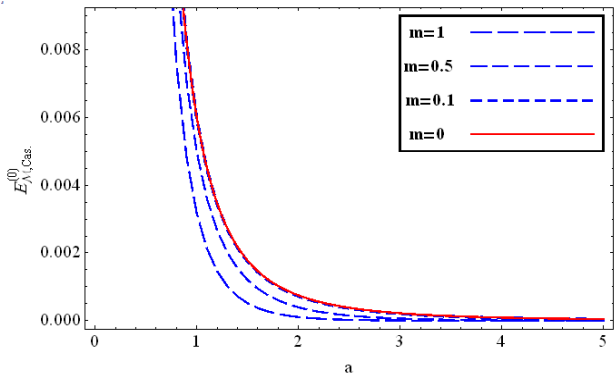

شکل ٢. در این نمودار میزان انرژی کازیمیر به ازاء جرمهای $m = \{1; 0.5; 0.1; 0\}$ بر حسب فاصله دو صفحه ($a$) رسم شده است. نزدیک شدن نمودار انرژی مربوط به میدان جرمدار به نمودار انرژی کازیمیر مربوط به میدان بدون جرم در روند کاهش جرم مشهود است.

توضیح اینکه مقادیر بدست آمده برای این مرتبه از انرژی کازیمیر با آنچه در مقالات گذشته در این خصوص گزارش شده است، تطابق کامل دارد[١١]. در شکل (٢)، مقادیر مربوط به انرژی کازیمیر را برای میدان اسکالر جرمدار با شرط مرزی مخلوط(دیریکله-نیومن) برای یک دسته نزولی از جرمهای مختلف رسم کرده‌ایم. روند نمودارهای رسم شده نشان می دهد که هر چه اندازه جرم کوچکتر می شود، نمودار مربوط به رابطه انرژی کازیمیرِ میدانِ جرمدار، به نمودار مربوط به میدان بدون جرم نزدیکتر می شود. در واقع این امر، سازگاری مناسب

---

پاسخ‌های بدست آمده را با مبانی قابل انتظار فیزیکی تأیید می نماید.

## تصحیح تابشی انرژی کازیمیر

لاگرانژی کلاین– گوردون برای میدان اسکالر جرمدار با جمله تراکنشی $\lambda\phi^4$ پس از اجرای برنامه بازبهنجارش بصورت زیر می باشد:

$$\mathcal{L} = \frac{1}{2}(\partial_\mu\phi)^2 - \frac{1}{2}m^2\phi^2 - \frac{\lambda}{4!}\phi^4 + \frac{1}{2}\delta_Z(\partial_\mu\phi)^2 - \frac{1}{2}\delta_m\phi^2 - \frac{\delta_\lambda}{4!}\phi^4$$

(۹)

که در آن $m$ و $\lambda$ پارامترهای فیزیکی جرم و ثابت جفت‌شدگی هستند و $\delta_Z$ و $\delta_m$ و $\delta_\lambda$ به ترتیب کانترترم‌های میدان، جرم و ثابت جفت شدگی می باشند. همانطور که می دانیم کانترترم‌ها موظف به هضم واگرایی‌های ناشی از پارامترهای برهنه در لاگرانژی و ارائه مقدار صحیح و فیزیکی این پارامترها هستند. اما نحوه انتخاب این کانترترم‌ها در محاسبات بازبهنجارش همیشه منشأ بسیاری از چالش‌ها بوده است. در مسائلی که میدان کوانتومی با یک شرط مرزی محدود می شود، این انتظار وجود دارد که کانترترم‌ها نیز با این شرط مرزی هماهنگ گردند. بنابراین استفاده از کانترترمِ مربوط به فضای آزاد در همه مسائل، بدون توجه به شرایط مرزی حاکم بر میدان، نمی تواند درست تلقی گردد. با توجه به اینکه کانترترم‌ها مسئول حذف واگرایی‌های ناشی از پارامترهای برهنه در لاگرانژی هستند، عدم انتخاب صحیح آن‌ها ممکن است به واگرا شدن کمیت‌های فیزیکی مرتبط با لاگرانژی در مسائل منجر گردد [۱۱]. در این مقاله، با در نظر گرفتن این ایده، از یک روش سیستماتیک و البته آسان برای بازبهنجارش پارامترهای برهنه استفاده شده است. این روش بازبهنجارش اجازه می دهد تا تاثیرات مرتبط با شرط مرزی در کانترترم انتخابی ورود یابد و البته این امر منجر به دستیابیِ ما به کانترترم وابسته به مکان نیز شده است. با استفاده از لاگرانژی

نوشته شده در معادله (۹) و بسط اختلالی می‌توان تابع انتشار را با استفاده از الگوهای فاینمن[1] به صورت زیر نوشت:

$$\underset{x_1 \qquad x_2}{\bullet\!-\!\bigcirc\!-\!\bullet} = \underset{x_1 \qquad x_2}{\bullet\!-\!\bullet} + \underset{x_1\ x\ \ x_2}{\bullet\!-\!\!\overset{\bigcirc}{\bullet}\!-\!\bullet} + \underset{x_1\ x\quad x_2}{\bullet\!-\!\otimes\!-\!\bullet} + \cdots ,$$

(۱۰)

که در آن $\underset{x_1\ x\ \ x_2}{\bullet\!-\!\otimes\!-\!\bullet}$ نشاندهنده[2] از کانترترم جرم میدان است. با استفاده شرط بازبهنجارش و اعمال آن تا مرتبه اول از $\lambda$ می توان مقادیر زیر را برای کانترترم نوشت. لذا داریم:

$$\delta_Z = 0 \quad , \quad \delta_\lambda = 0$$

$$\delta_m = \frac{-i}{2}\ \underset{x}{\bigcirc} = \frac{-\lambda}{2}G(x,x)$$

(۱۱)

که در آن $G(x,x')$ تابع گرین(انتشار[3]) می باشد. حال با عنایت به تعریف رابطه انرژی خلا تا مرتبه اول ثابت جفت‌شدگی $\lambda$ بصورت زیر، و استفاده از مقدار بدست آمده برای کانترترم جرم از رابطه (۱۱)، داریم: [۱۲-۱۵]

$$E^{(1)} = i\int_V \left(\frac{1}{8}\bigcirc\!\bigcirc + \frac{1}{2}\otimes\right)d^3x$$

$$= i\int_V \left[\frac{-i\lambda}{8}G^2(x,x) - \frac{i}{2}\delta_m(x)G(x,x)\right]d^3x$$

$$= \frac{-\lambda}{8}\int_V G^2(x,x)d^3x$$

(۱۲)

مقدار تابع گرین را می توان برای میدان اسکالر تعریف شده در لاگرانژی رابطه (۹) بین دو صفحه موازی که محدود کننده میدان با شرط مرزی مخلوط (دیریکله-نیومن) است پس از انجام چرخش ویک[4] بصورت زیر نوشت:

---

1 Feynman diagram
2 Symbollicaly

3 The propagator
4 Wick's rotation

هـر انتگـرال را بـه وضـوح تشخیص داد. بـا تکـرار کـل ایـن روند بـرای همـه جمـلات مشـابه موجـود در انـرژی خـلا مربوط به همه نـواحی تعریـف شـده در شکـل (۱)، مـی‌تـوان رابطه اختلاف دو انرژی خلا را بصورت زیر نوشت:

$$
\begin{aligned}
E_A^{(1)} - E_B^{(1)} &= \frac{-\lambda L^2}{16\pi^4 a}\Bigg[\left(\sum_{n=0}^{\infty}\left(\Lambda - \frac{\pi}{2}\omega_{n.A1} + \frac{\omega_{n.A1}^2}{\Lambda}\right)\right)^2 \\
&\quad + \frac{1}{2}\sum_{n=0}^{\infty}\left(\Lambda - \frac{\pi}{2}\omega_{n.A1} + \frac{\omega_{n.A1}^2}{\Lambda}\right)^2\Bigg] \\
&\quad + \frac{-\lambda L^2}{4\pi^4(L-a)}\Bigg[\left(\sum_{n=0}^{\infty}\left(\Lambda' - \frac{\pi}{2}\omega_{n.A2}\right.\right. \\
&\quad \left.\left. + \frac{\omega_{n.A2}^2}{\Lambda'}\right)\right)^2 \\
&\quad + \frac{1}{2}\sum_{n=0}^{\infty}\left(\Lambda' - \frac{\pi}{2}\omega_{n.A2} + \frac{\omega_{n.A2}^2}{\Lambda'}\right)^2\Bigg] \\
&\quad - \{a \to b\}
\end{aligned}
$$

$$(۱۵)$$

با انتخاب مقادیر مناسب از مقدارهای قطع می‌توان نشان داد که سهم کلیه جملات واگرای ناشی از انتگرال در عبارات فوق یکدیگر را حذف کرده و جملات باقیمانده بصورت زیر می‌باشد:

$$
\begin{aligned}
E_A^{(1)} - E_B^{(1)} &= \frac{-\lambda L^2}{64\pi^2 a}\Bigg[\left(\sum_{n=0}^{\infty}\omega_{n.A1}\right)^2 \\
&\quad + \frac{1}{2}\left(1 + \frac{8}{\pi^2}\right)\sum_{n=0}^{\infty}\omega_{n.A1}^2\Bigg] \\
&\quad + \frac{-\lambda L^2}{16\pi^2(L-a)}\Bigg[\left(\sum_{n=0}^{\infty}\omega_{n.A2}\right)^2 \\
&\quad + \frac{1}{2}\left(1 + \frac{8}{\pi^2}\right)\sum_{n=0}^{\infty}\omega_{n.A2}^2\Bigg] - \{a \to b\}
\end{aligned}
$$

$$(۱۶)$$

عبارات جمع موجود در رابطه (۱۶) همچنان واگرا هستند و برای اینکه بتوان آن‌ها را منظم نمود، معمولاً بهترین روش تبدیل آنان به

$$
\begin{aligned}
&G_{A1}(x,x') \\
&= \frac{1}{a}\int \frac{d^3\mathbf{k}}{(2\pi)^3}\sum_{n=0}^{\infty}e^{-\omega(t-t')}e^{-ik_x(x-x')}e^{-ik_y(y-y')}\times
\end{aligned}
$$

$$
\frac{[\sin k_n z + (-1)^n \cos k_n z][\sin k_n z' + (-1)^n \cos k_n z']}{k^2 + k_n^2 + m^2}
$$

$$(۱۳)$$

که در آن $k^2 = \omega^2 + k_x^2 + k_y^2$ و $k_n = \frac{\pi}{a}\left(n + \frac{1}{2}\right)$ و $m$ جرم میدان اسکالر می باشد. پارامتر $a$ نیز فاصله بین دو صفحه موازی در ناحیه $A1$ شکـل (۱) می‌باشـد. منظور از $d^3\mathbf{k}$ نیز عبـارت $d^3\mathbf{k} = d\omega dk_x dk_y$ ا است. با جایگذاری تابع گرین در رابطه (۱۲)، می‌توان فرم انرژی خلا را  تا مرتبه اول $\lambda$ بصورت زیر نوشت:

$$
\begin{aligned}
&E_{A1}^{(1)} \\
&= \frac{-\lambda L^2}{4a}\sum_{n.n'=0}^{\infty}\left(\int \frac{d^3\mathbf{k}}{(2\pi)^3}\frac{1}{k^2 + \omega_{n.A1}^2}\right) \\
&\quad \times \left(\int \frac{d^3\mathbf{k}}{(2\pi)^3}\frac{1}{k^2 + \omega_{n'.A1}^2}\right)\left(1 + \frac{1}{2}\delta_{nn'}\right) \\
&= \frac{-\lambda L^2}{4a}\Bigg[\left(\sum_{n=0}^{\infty}\int_0^{\infty}\frac{4\pi k^2 dk}{(2\pi)^3}\frac{1}{k^2 + \omega_{n.A1}^2}\right)^2 \\
&\quad + \frac{1}{2}\sum_{n=0}^{\infty}\left(\int_0^{\infty}\frac{4\pi k^2 dk}{(2\pi)^3}\frac{1}{k^2 + \omega_{n'.A1}^2}\right)^2\Bigg]
\end{aligned}
$$

$$(۱۴)$$

کـه در آن $\omega_{n.A1} = \sqrt{k_n^2 + m^2}$ مـی‌باشـد. هـردو انتگـرال موجود در کروشه در معادله فوق واگرا هستند. فلـذا، بـرای منظم‌سازی و حـذف چنیـن واگرایـی‌هـایی نیاز بـه اسـتفاده از یـک روش مـنظم‌سـازی اسـت. بدیـن منظـور از مـنظم سـازی کـم کردن جعبـه هـا و مـنظم سازی تـابع قطع[1] بـصـورت همزمـان بهـره مـی جـوئیم. لـذا حـد بـالای انتگرال‌هـا را در معادله فوق با یـک مقـدار قطـع بـه ماننـد $\Lambda$ عـوض مـی کنیم و سـپس انتگرال‌هـا را محاسـبه کـرده و پاسـخ انتگرال‌ها را کـه تـابعی از مقـدار قطـع $\Lambda$ اسـت در حـد $\Lambda \to \infty$ بـسط مـی‌-دهیم. این امر باعـث مـی‌شـود تـا بتـوان تکـه‌هـای واگراکننـده

---

[1] Cutoff regularization technique

$$\frac{-\lambda L^2 m^4}{128\pi^2}\left(1+\frac{8}{\pi^2}\right)\left[\frac{1}{a}\mathcal{I}_2(a)+\frac{4}{L-a}\mathcal{I}_2\left(\frac{L-a}{2}\right)-\{a\to b\}\right]$$
$$=\frac{-\lambda L^2 m^3}{128\pi^3}\left(1+\frac{8}{\pi^2}\right)\left[a+2\frac{L-a}{2}-b\right.$$
$$\left.-2\frac{L-b}{2}\right]\int_0^\infty(\xi^2+1)d\xi=0$$

$$(۱۹)$$

با توجه به رابطهٔ (۱۹) و پس از بسط جملهٔ مربع در رابطهٔ (۱۷) و با در نظر گرفتن اینکه اندازهٔ $\mathcal{B}_2(a)$ برابر صفر است، عبارت (۱۷) بصورت زیر نوشته خواهد شد:

$$E_A^{(1)}-E_B^{(1)}=\frac{-\lambda L^2 m^4}{64\pi^2 a}[\mathcal{I}_1(a)^2+2\mathcal{I}_1(a)\mathcal{B}_1(a)$$
$$+\mathcal{B}_1(a)^2]$$
$$+\frac{-\lambda L^2 m^2}{16\pi^2(L-a)}\left[\mathcal{I}_1\left(\frac{L-a}{2}\right)^2\right.$$
$$+2\mathcal{I}_1\left(\frac{L-a}{2}\right)\mathcal{B}_1\left(\frac{L-a}{2}\right)+\mathcal{B}_1\left(\frac{L-a}{2}\right)^2\right]$$
$$-\{a\to b\},$$

$$(۲۰)$$

روشن است که جملهٔ اول کروشهها در رابطهٔ فوق نیز واگراست. اما می توان با توجه به فرایند تعریف شده در روش کم کردن جعبه ها این جملات واگرا را بصورت کامل حذف نمود. لذا، برای جملهٔ اول کروشهها در معادلهٔ (۲۰) داریم:

$$\frac{-\lambda L^2 m^4}{64\pi^2}\left[\frac{1}{a}\mathcal{I}_1(a)^2+\frac{4}{L-a}\mathcal{I}_1\left(\frac{L-a}{2}\right)^2-\{a\to b\}\right]$$
$$=\frac{-\lambda L^2 m^4}{64\pi^4}\left[a+2\frac{L-a}{2}-b\right.$$
$$\left.-2\frac{L-b}{2}\right]\left(\int_0^\infty\sqrt{\xi^2+1}\,d\xi\right)^2=0$$

$$(۲۱)$$

جملهٔ دوم در کروشه های معادلهٔ (۲۰) نیز بدلیل حضور جملهٔ $\mathcal{I}_1(x)$ واگراست. برای حذف این واگرایی علاوه بر روش کم کردن جعبه ها، از منظم سازی تابع قطع نیز استفاده می کنیم. بدین منظور حد بالای انتگرال $\mathcal{I}_1(a)$ را با یک مقدار قطع $\Lambda$ عوض می کنیم. پس از محاسبهٔ انتگرال، پاسخ را در حد $\infty\to\Lambda$ بسط خواهیم داد. اکنون می توان با تکرار این امر برای کلیه جملات مشابه دیگر در کروشه های معادلهٔ (۲۰)، بخش واگراکننده را

انتگرال است. لذا با استفاده از رابطه ایبل-پلانا که در معادله (۳) معرفی شده است، داریم:

$$E_A^{(1)}-E_B^{(1)}$$
$$=\frac{-\lambda L^2 m^4}{64\pi^2 a}\left[\left(\underbrace{\frac{a}{\pi}\int_0^\infty\sqrt{\xi^2+1}\,d\xi}_{\mathcal{I}_1(a)}+\mathcal{B}_1(a)\right)^2\right.$$
$$+\frac{1}{2}\left(1+\frac{8}{\pi^2}\right)\left(\underbrace{\frac{a}{m\pi}\int_0^\infty(\xi^2+1)d\xi}_{\mathcal{I}_2(a)}+\mathcal{B}_2(a)\right)\right]$$
$$+\frac{-\lambda L^2 m^4}{16\pi^2(L-a)}\left[\left(\underbrace{\frac{L-a}{2\pi}\int_0^\infty\sqrt{\xi^2+1}\,d\xi}_{\mathcal{I}_1\left(\frac{L-a}{2}\right)}+\mathcal{B}_1\left(\frac{L-a}{2}\right)\right)^2\right.$$
$$\left.+\frac{1}{2}\left(1+\frac{8}{\pi^2}\right)\left(\underbrace{\frac{(L-a)}{2m\pi}\int_0^\infty(\xi^2+1)d\xi}_{\mathcal{I}_2\left(\frac{L-a}{2}\right)}+\mathcal{B}_2\left(\frac{L-a}{2}\right)\right)\right]$$
$$-\{a\to b\}$$

$$(۱۷)$$

که در آن $\mathcal{B}_1(a)$ و $\mathcal{B}_2(a)$ مقادیر برنچکات رابطه ایبل-پلانا بوده و مقادیر آنان بصورت زیر نوشته می شود:

$$\mathcal{B}_1(a)=\frac{2a}{\pi}\int_1^\infty\frac{\sqrt{\eta^2-1}}{e^{2ma\eta}+1}d\eta$$
$$=\frac{2a}{\pi}\sum_{j=1}^\infty\int_1^\infty(-1)^{j+1}e^{-2ma\eta j}\sqrt{\eta^2-1}\,d\eta$$
$$=\frac{1}{m\pi}\sum_{j=1}^\infty\frac{(-1)^{j+1}K_1(2maj)}{j},$$
$$\mathcal{B}_2(a)=0$$

$$(۱۸)$$

جملهٔ $\mathcal{I}_1(x)$ و $\mathcal{I}_2(x)$ در رابطه (۱۷) هردو جمله انتگرالی رابطه ایبل-پلانا بوده و واگرا هستند، و حذف واگرایی آنها با استفاده از طرح کم کردن جعبه ها براحتی امکانپذیر است. برای جمله $\mathcal{I}_2(x)$ داریم:

بصورت شفاف پیدا کرد. انتخاب مناسب از مقادیر قطع کمک می کند تا سهم واگرا را از جملهٔ بسط حذف کرده و آنچه از جمله دوم کروشه‌های معادله (۲۰) باقی می‌ماند به فرم زیر می‌باشد[1]:

$$
\begin{aligned}
\frac{-\lambda L^2 m^4}{64\pi^3}\bigg[ & \left(\Lambda^2 + \frac{1+2\ln 2}{2} + \ln\Lambda + \mathcal{O}(\Lambda^{-2})\right)\mathcal{B}_1(a) \\
& + 2\left(\Lambda'^2 + \frac{1+2\ln 2}{2} + \ln\Lambda' \right. \\
& \left. + \mathcal{O}(\Lambda'^{-2})\right)\mathcal{B}_1(\tfrac{L-a}{2}) - \{a \to b\}\bigg] \\
& = \frac{-\lambda L^2 m^4}{128\pi^3}(1+2\ln 2)\big[\mathcal{B}_1(a) \\
& + 2\mathcal{B}_1(\tfrac{L-a}{2}) - \{a \to b\}\big].
\end{aligned}
$$

(۲۲)

لذا آنچه از اختلاف انرژی‌های خلا در معادله (۲۰) باقی می‌ماند بصورت زیر است:

$$
\begin{aligned}
E_A^{(1)} & - E_B^{(1)} \\
= & \frac{-\lambda L^2 m^4}{64\pi^2 a}\bigg[\frac{a(1+2\ln 2)}{2\pi}\mathcal{B}_1(a) + \mathcal{B}_1(a)^2\bigg] \\
& + \frac{-\lambda L^2 m^4}{16\pi^2(L-a)}\bigg[\frac{(L-a)(1+2\ln 2)}{4\pi}\mathcal{B}_1(\tfrac{L-a}{2}) \\
& + \mathcal{B}_1(\tfrac{L-a}{2})^2\bigg] - \{a \to b\}
\end{aligned}
$$

(۲۳)

برای بدست آوردن انرژی کازیمیر لازم است، آخرین گام از تعریف ارائه شده در معادله (۱) اجرا شود. بدین منظور به محاسبه حدهای $\infty \to b$ و $\infty \to L$ می‌پردازیم. رابطه تصحیح تابشی انرژی کازیمیر برای میدان اسکالر جرمدار محدود بین دو صفحه موازی با فاصله $a$ از یکدیگر با شرط مرزی مخلوط (دیریکله-نیومن) بصورت زیر بدست خواهد آمد:

$$
E_{\text{Cas.}}^{(1)} = \frac{-\lambda L^2 m^4}{64\pi^3}\bigg[\frac{(1+2\ln 2)}{2}\mathcal{B}_1(a) + \frac{\pi}{a}\mathcal{B}_1(a)^2\bigg].
$$

(۲۴)

برای میدان بدون جرم و میدان با جرم سنگین رابطه انرژی کازیمیر بدست آمده در رابطه فوق به فرم زیر تبدیل می شود:

$$
E_{\text{Cas.}}^{(1)} \to \begin{cases} \dfrac{-\lambda L^2}{36864 a^3}, & m = 0; \\[3mm] \dfrac{-\lambda L^2}{256\pi^3 a^3}(1+2\ln 2)\left(\dfrac{ma}{\pi}\right)^{\frac{5}{2}}e^{-2ma}, & m \gg 1. \end{cases}
$$

(۲۵)

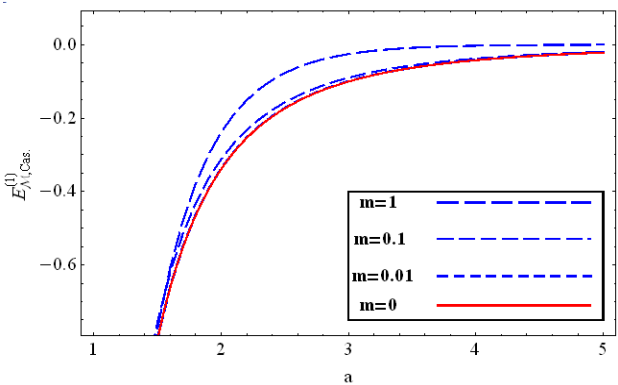

شکل ۳. در این نمودار، میزان انرژی کازیمیر برای میدان اسکالر بین دو صفحه موازی با شرط مرزی مخلوط(دیریکله-نیومن) بر حسب فاصله صفحات ($a$) رسم شده است. این نمودار نشان می دهد که با کاهش اندازه جرم، نمودار انرژی برای میدان اسکالر جرمدار به سرعت به نمودار انرژی برای میدان بدون جرم نزدیک می‌شود و تفاوت قابل ملاحظه ای در جرم‌های کمتر از ۰٫۰۱ مشاهده نمی‌شود.

توضیح اینکه پاسخهای بدست آمده با آنچه در [۱۱] گزارش شده متفاوت بوده و ریشه این تفاوت را می‌توان در انتخاب نوع بازبهنجارش جستجو نمود. البته پاسخ بدست آمده با مبانی فیزیکی سازگار است و همانطور که از روند محاسبات برمی آید، انجام فرایند محاسبه بدون هیچگونه ابهام و یا پیچیدگی خاصی صورت پذیرفته است. در نمودار شکل (۳)، تصحیح تابشی انرژی کازیمیر برحسب تابعی از فاصله صفحات $a$ به ازای چند جرم مختلف میدان رسم شده است. روند نمودارهای رسم شده نشان می دهد که با کاهش میزان جرم، نمودار انرژی مربوط به میدان

رابطه (۲۲) باقی نماند. توضیح اینکه درجه آزادی کافی برای انتخاب مناسب مقادیر قطع $\Lambda_{A1}$، $\Lambda_{A2}$، $\Lambda_{B1}$ و $\Lambda_{B2}$ وجود دارد و می‌توان گفت عملاً این درجات آزادی، از نوع تکنیک بکار گرفته شده و ساختارهای انتخاب شده در روش کم کردن جعبه‌ها حاصل شده است.

همانطور که می دانیم، بین روابط انرژی کازمیر برای میدان اسکالر بین دو صفحه موازی در ۳+۱ بعد با شرایط مرزی دیریکله، نیومن و نوسانی روابطی به شکل زیر برقرار است:

$$E_{\mathcal{P}}(a) = 2E_{\mathcal{D}}\left(\frac{a}{2}\right), \qquad E_{\mathcal{N}}(a) = E_{\mathcal{D}}(a)$$

(۲۶)

توضیح اینکه انرژی کازمیر برای میدان اسکالر با شرط مرزی دیریکله در گذشته محاسبه شده است [۱۱] و این روابط کمک می کند تا بتوان انرژی کازمیر را برای شرط مرزی نیومن و نوسانی از روی روابط انرژی مربوط به شرط مرزی دیریکله بازنویسی نمائیم. در نمودار شکل (۵)، چگالی انرژی کازمیر را برای چهار شرط مرزی دیریکله، نیومن، مخلوط آنان (دیریکله – نیومن) و نوسانی را در کنار هم رسم کرده‌ایم. این نمودارها کمک می کنند تا علامت انرژی‌ها و همچنین مرتبه بزرگی آنها در کنار هم دیده شوند. همانطور که مشخص است علامت این انرژی برای شرط مرزی مخلوط (دیریکله-نیومن) در مرتبه صفرم مثبت و برای دیگر شرایط مرزی منفی می باشد.

## بحث و نتیجه‌گیری

در این مقاله، به محاسبه مرتبه صفرم و اول تصحیح تابشی انرژی کازمیر برای نظریه $\phi^4$ بین دو صفحه موازی در ۳+۱ بعد با شرط مرزی مخلوط (دیریکله-نیومن) پرداخته شده است. دو نکته در انجام این محاسبات بسیار حائز اهمیت است. یکی اینکه، برای بازبهنجارش پارامترهای برهنه در این محاسبات از یک روش سیستماتیک بهره جسته ایم که در آن تمامی المانهای موجود در پروسه بازبهنجارش سازگار با شرط مرزی حاکم بر مساله هستند. این امر موجب ظهور کانترترم‌های وابسته به مکان شده است. استفاده از روش منظم سازی کم کردن جعبه‌ها، از دیگر نکات حائز اهمیت در محاسبات صورت گرفته در این مقاله می باشد. مزیت و برتری این روش منظم سازی، در عدم استفاده از تمدید تحلیلی و سادگی حذف واگرایی‌ها در پروسه انجام این است. پاسخ های بدست آمده با آنچه در گذشته در خصوص این مساله در مرتبه اول تصحیح تابشی انرژی کازمیر گزارش شده، متفاوت است. این تفاوت را می توان در انتخاب برنامه

جرمدار به نمودار انرژی مربوط به میدان بدون جرم نزدیک می شود، که این امر با انتظارات فیزیکی سازگار است. در نمودار شکل (۴)، مرتبه صفرم و اول تصحیح تابشی انرژی کازمیر را برای میدان بدون جرم و جرمدار در کنار هم رسم نموده‌ایم. این نمودار نشان می دهد به ازای مقدار ثابت جفت شدگی $\lambda = 0.1$ ، میزان انرژی کازمیر در مرتبه اول تصحیح تابشی تقریباً هزار برابر کوچکتر از این انرژی در مرتبه صفرم است.

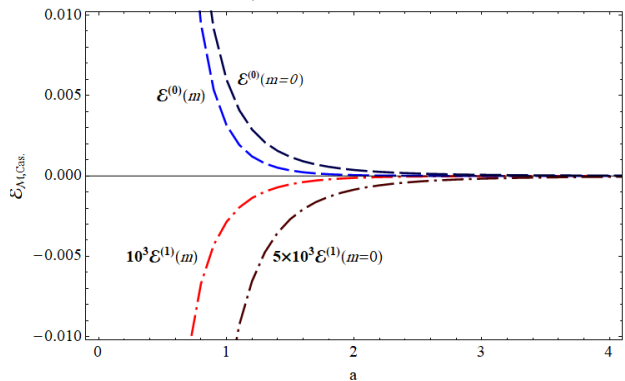

شکل ۴. در این نمودار میزان چگالی انرژی کازمیر در مرتبه صفرم و اول تصحیح تابشی به ازاء جرم $m = \{1; 0\}$ برحسب فاصله صفحات $(a)$ رسم شده است. این نمودار براحتی نشان می دهد که مرتبه اول تصحیح تابشی انرژی کازمیر برای میدان اسکالر محدود شده با شرط مرزی مخلوط دارای مقدار منفی بوده و اندازه آن حدوداً هزار برابر کوچکتر از مرتبه صفرم است. مقدار ثابت جفت شدگی در نمودارها برابر $\lambda = 0.1$ در نظر گرفته شده است.

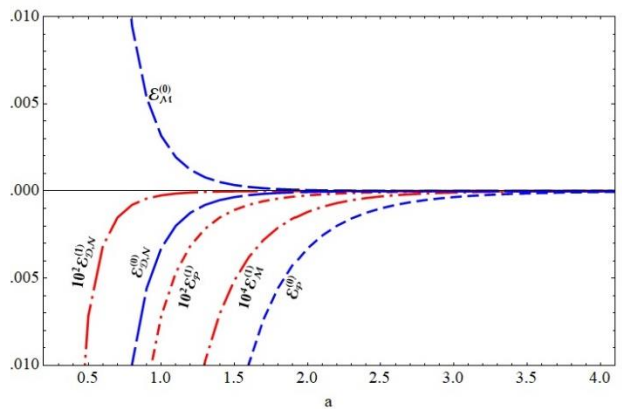

شکل ۵. در این نمودار مرتبه های صفرم و اول چگالی انرژی کازیمیر مربوط به میدان اسکالر محدود شده بین دو صفحه موازی با فاصله صفحات $a$ با شرط مرزی دیریکله، نیومن، مخلوط(دیریکله-نیومن) و نوسانی بر حسب فاصله صفحات $(a)$ رسم شده است. با توجه به این نمودار می توان علامت انرژی و اندازه های آن را برای شروط مرزی مختلف در کنار هم مقایسه نمود. در کلیه این نمودارها، اندازه ثابت جفت‌شدگی و جرم میدان به ترتیب $\lambda = 0.1$ و $m = 1$ در نظر گرفته شده است.

بازبهنجارش جستجو کرد. در این مقاله همچنین، مقادیر انرژی کازیمیر مربوط به شرایط مرزی دیریکله، نیومن، مخلوط(دیریکله-نیومن) و نوسانی در کنار هم مقایسه شده اند. این مقایسه نشان می دهد که، مقادیر مربوط به این انرژی ها در مرتبه اول تصحیح تابشی برای کلیه شرایط مرزی، منفی است و اندازه آنها نیز از اندازه مرتبه صفرم مربوطه کوچکتر است. البته کلیه مقادیر بدست آمده درمرتبه اول تصحیح تابشی با آنچه در

گذشته در این خصوص گزارش شده است، تفاوت دارد؛ که علت اصلی این تفاوت را باید برنامه بازبهنجارش استفاده شده دانست.

# مراجع

# Radiative Correction to The Casimir Energy For Scalar Field with Mixed Boundary Condition in 3 + 1 Dimensions

## M. A. Valuyan[1]

Department of Physics, Semnan Branch, Islamic Azad University, Semnan, Iran

## Abstract

In the present study, the zero and first-order radiative correction to the Casimir energy for massive and massless scalar fields confined with mixed boundary conditions (Dirichlet-Neumann) between two parallel plates in $\phi^4$ theory were computed. Two issues in performing the calculations in this work are essential: to renormalize the bare parameters of the problem, a systematic method were used, which allowing all influences from the boundary conditions to be imported in all elements of the renormalization program. This idea yields our counterterms appeared in the renormalization program to be position-dependent. Using the Box Subtraction Scheme as a regularization technique is the other noteworthy point in the calculation. In this scheme, by subtracting the vacuum energies of two similar configurations from each other, regularizing divergent expressions and their removal process were significantly facilitated. All the obtained answers for the Casimir energy with the mixed boundary condition were consistent with well-known physical grounds. We also compared the Casimir energy for massive scalar field confined with four types of boundary conditions (Dirichlet, Neumann, mixed of them and Periodic) in 3+1 dimensions with each other, and the sign and magnitude of their values were discussed.

**Keywords:** Casimir Energy, Renormalization, Boundary Condition, Regularization

---

[1] Corresponding author: m-valuyan@sbu.ac.ir